\documentclass[10pt,conference]{IEEEtran}
\IEEEoverridecommandlockouts
\usepackage[T1]{fontenc}
\usepackage{amsmath}
\usepackage{amsfonts}
\usepackage{braket}
\usepackage{tikz}
\usetikzlibrary{arrows.meta, positioning, fit, shapes.geometric}
\usepackage{quantikz}
\usepackage{enumitem}
\usepackage[acronym,nohypertypes={acronym},shortcuts]{glossaries}
\usepackage[caption=false]{subfig}
\usepackage{multirow}
\usepackage[hyphens]{url}
\usepackage[hidelinks]{hyperref}
\usepackage{graphicx}
\usepackage{color}

\tikzset{
    every node/.style={font=\normalsize}
}
\usepackage{cite}
\usepackage{amsmath,amssymb,amsfonts}
\usepackage{algorithmic}
\usepackage{graphicx}
\usepackage{textcomp}
\usepackage{xcolor}
\def\BibTeX{{\rm B\kern-.05em{\sc i\kern-.025em b}\kern-.08em
    T\kern-.1667em\lower.7ex\hbox{E}\kern-.125emX}}
\begin{document}
\title{Strengthening security and noise resistance in one-way quantum key distribution protocols through hypercube-based quantum walks}
 
\author{
\IEEEauthorblockN{David Polzoni}
\IEEEauthorblockA{\textit{University of Padova}\\
Padova, Italy \\
david.polzoni@studenti.unipd.it \\
\textit{Institute of Physics Slovak Academy of Sciences} \\
Bratislava, Slovakia \\
david.polzoni@savba.sk}
\and
\IEEEauthorblockN{Tommaso Bianchi}
\IEEEauthorblockA{\textit{University of Padova}\\
Padova, Italy \\
tommaso.bianchi@phd.unipd.it}
\and
\IEEEauthorblockN{Mauro Conti}
\IEEEauthorblockA{\textit{University of Padova}\\
Padova, Italy \\
mauro.conti@unipd.it \\
\textit{Örebro University}\\
Örebro, Sweden}
}
%---------------------------------------------------------------------
%%% Acronyms
%---------------------------------------------------------------------
\newacronym{qkd}{QKD}{Quantum Key Distribution}
\newacronym{qws}{QWs}{Quantum Walks}
\newacronym{qw}{QW}{Quantum Walk}
\newacronym{qer}{QER}{Quantum Error Rate}
\newacronym{vm}{VM}{Virtual Machine}
%---------------------------------------------------------------------
 
\maketitle

%---------------------------------------------------------------------

\begin{abstract}
\ac{qkd} is a foundational cryptographic protocol that ensures information-theoretic security.
However, classical protocols such as BB84, though favored for their simplicity, offer limited resistance to eavesdropping, and perform poorly under realistic noise conditions.
Recent research has explored the use of discrete-time \ac{qws} to enhance \ac{qkd} schemes.
In this work, we specifically focus on a one-way \ac{qkd} protocol, where security depends exclusively on the underlying \ac{qw} topology, rather than the details of the protocol itself. Our paper introduces a novel protocol based on \ac{qws} over a hypercube topology and demonstrates that, under identical parameters, it provides significantly enhanced security and noise resistance compared to the circular topology (i.e., state-of-the-art), thereby strengthening protection against eavesdropping.
Furthermore, we introduce an efficient and extensible simulation framework for one-way \ac{qkd} protocols based on \ac{qws}, supporting both circular and hypercube topologies.
Implemented with IBM’s software development kit for quantum computing (i.e., Qiskit), our toolkit enables noise-aware analysis under realistic noise models.
To support reproducibility and future developments, we release our entire simulation framework as open-source.
This contribution establishes a foundation for the design of topology-aware \ac{qkd} protocols that combine enhanced noise tolerance with topologically driven security.
\end{abstract}

\begin{IEEEkeywords}
Quantum Key Distribution, Quantum Walks, Hypercubes.
\end{IEEEkeywords}

%---------------------------------------------------------------------
\section{Introduction}\label{sec:introduction}
%---------------------------------------------------------------------

\noindent
\acrfull{qkd} is a core application of quantum cryptography, allowing two parties to establish a shared secret key with security guaranteed by quantum mechanics, since any eavesdropping attempt that disturbs the system becomes detectable. 
The first protocols, BB84~\cite{bennett_bb84} and E91~\cite{ekert_e91}, laid the foundation for modern implementations, which now include commercial systems, long-distance links, and even satellite-based communication~\cite{qkd_satellites}. 
\acrfull{qws}, introduced by Aharonov et al.~\cite{quantum_random_walks}, extend classical random walks into the quantum domain and are central to quantum computing. 
Their cryptographic applications are emerging, such as the protocol of Rohde et al.~\cite{rohde_qkd_quantum_walks}, which enables quantum homomorphic encryption, allowing servers to process encrypted quantum data while preserving privacy.

\noindent
\textbf{Contributions}. Our work builds on Vlachou et al.~\cite{vlachou_qkd_quantum_walks}, who used \acrfull{qw} properties to design secure \ac{qkd} protocols, including a one-way (prepare-and-measure) scheme with a security proof against full man-in-the-middle attacks on circle graphs (i.e., the current state of the art).
We generalize this security proof to arbitrary regular topologies and introduce a hypercube-based \ac{qws} protocol.
We also develop a realistic, reproducible Qiskit model for simulating \ac{qws} on circle and hypercube topologies and use it to implement the protocol.
Simulations under Qiskit's built-in noise models show that the hypercube-based approach outperforms the state of the art in both security and noise resilience.
Our main contributions are:

\begin{itemize}[label=\textbullet]
    \item We introduce a novel one-way \ac{qkd} protocol based on discrete-time \ac{qws} over hypercube topologies, offering a new direction beyond the circular structures;
    \item We develop robust, flexible, and extensible Qiskit-based models to simulate discrete-time \ac{qws} on both circular and hypercube topologies, enabling systematic evaluation across graph structures;
    \item We extend prior findings that the security of \ac{qw}-based \ac{qkd} protocols is determined solely by the underlying walk structure, generalizing this dependence to arbitrary topologies, and demonstrate that hypercube-based \ac{qws} achieve higher security by providing increased resistance to eavesdropping;
    \item We implement and compare a one-way \ac{qkd} protocol over both \ac{qws} models, introducing a unified simulation framework for such a comparative study within the Qiskit environment;
    \item We show that the hypercube-based protocol improves the maximally tolerated \acrfull{qer} under depolarizing noise by approximately $20\%$, and by $13\%$ under combined amplitude-phase damping, compared to the circular case, under identical conditions.
\end{itemize}

\noindent
To support reproducibility and further research, we have made the source code publicly available on GitHub~\cite{github_repo}.

\noindent
\textbf{Organization}. Section~\ref{sec:related_works} reviews related work and Section~\ref{sec:background} covers background; Section~\ref{sec:qkd_protocol} presents the protocol and its security proof; Section~\ref{sec:qiskit_model} details the Qiskit implementation; Sections~\ref{sec:security_evaluation} and~\ref{sec:noise_resistance_analysis} evaluate security and noise resilience; and Sections~\ref{sec:discussion} and~\ref{sec:conclusions} discuss limitations and conclude.

%---------------------------------------------------------------------
\section{Related works}\label{sec:related_works}
%---------------------------------------------------------------------

\noindent
This section reviews related works defining the state-of-the-art prior to our contribution.
The main reference for \ac{qkd} protocols based on \ac{qws} is Vlachou et al.~\cite{vlachou_qkd_quantum_walks}, who proposed a secure protocol with verification against full man-in-the-middle attacks.
They also present a one-way \ac{qkd} protocol, supported by a security proof from its entanglement-based version, and explore a semi-quantum variant.
The most advanced implementation of the one-way protocol uses a circular topology, analyzing noise resistance under a generalized Pauli channel and providing the maximally tolerated \ac{qer}, serving as the state-of-the-art reference.
Building on this, we propose a hypercube-based \ac{qws} protocol achieving improved robustness and a higher tolerated \ac{qer} under the same conditions.
For Qiskit-based \ac{qws} implementation, we follow Douglas et al.~\cite{douglas_wang_eff_circuit_quantum_walks}, who define discrete-time \ac{qw} on general undirected graphs.
Each vertex $v_i$ with degree $d_i$ is split into $d_i$ subnodes, and the shift operator $S(v_i, a_i) = (v_j, a_j)$ moves states along edges $(v_i, v_j) \in E$, while the coin operator mixes amplitudes among subnodes via a $d_i \times d_i$ unitary matrix.
For a circle, two subnodes per node are encoded with $\lceil\log_2(n)\rceil$ qubits plus one subnode qubit; the coin acts on the subnode qubit, and the shift operator performs cyclic permutations via \textsc{I} and \textsc{D} gates:

\[
    S = (\textsc{I} \otimes \ket{1} + \textsc{D} \otimes \ket{0}).
\]

\noindent
This framework scales efficiently to hypercube-based \ac{qws}, which require larger state and coin spaces, enabling our hypercube-based \ac{qkd} protocol, as detailed in the following sections.

%---------------------------------------------------------------------
\section{Background}\label{sec:background}
%---------------------------------------------------------------------

\noindent
In this section, we present the theoretical background relevant to our work. We begin by reviewing the fundamentals of quantum walks and the quantum gates involved. We then describe the formal models that govern quantum walks on both circular and hypercube structures.

%---------------------------------------------------------------------
\subsection{Quantum walks basics and quantum gates}\label{subsec:quantum_walks_gates}
%---------------------------------------------------------------------

\noindent
This paper compares circle and hypercube-based \ac{qws}.
Unlike classical random walks, where the next state depends only on the current one~\cite{classical_random_walks}, \ac{qws} explore all paths simultaneously via quantum superposition~\cite{quantum_walks_review}.
In coined \ac{qws}, the walker's position resides in an $n$-dimensional space with $2^n$ positions, encoded in $\mathcal{H} = \mathcal{H}_c \otimes \mathcal{H}_p$, where $\mathcal{H}_c = \{\ket{0}, \ket{1}\}$ is the coin space and $\mathcal{H}_p = \{\ket{j}: j = 0, \dots, n-1\}$ the position space.
Evolution is governed by coin and shift operators, producing the unitary operator:

\begin{equation}\label{eq:intro_unitary_qw}
    U = S \cdot (C \otimes I),
\end{equation}

\noindent
where $S$ updates positions based on the coin state, $C$ is a coin operator, and $I$ preserves the walker's state.
Some classical random walk principles can extend to \ac{qws}~\cite{quantum_walk_computing}, with the system after $t$ steps given by:

\begin{equation}\label{eq:q_state_after_t}
    \ket{\psi(t)} = U^t\ket{\psi(0)}.
\end{equation}

\noindent
To construct our \ac{qw} models, we employ the Pauli-$X$ (i.e., \textsc{NOT}), Hadamard ($H$), and phase ($S$) gates, which serve respectively to initialize the walker's position, create coin superpositions, and balance the evolution through phase shifts~\cite{quantum_computation_information}.

%---------------------------------------------------------------------
\subsection{Circle-based quantum walks}\label{subsec:circle_quantum_walks}
%---------------------------------------------------------------------

\noindent
In this context, following the modeling from Vlachou et al.~\cite{vlachou_qkd_quantum_walks}, the walker moves between discrete positions on a circle.
The Hilbert space $\mathcal{H}$, describing the \ac{qw}, is the tensor product $\mathcal{H} = \mathcal{H}_p \otimes \mathcal{H}_c$, where $\mathcal{H}_p$ is spanned by position states $\{\ket{x}: x \in {0, \dots, P - 1}\}$, with $P$ denoting the number of discrete positions, and $\mathcal{H}_c$ by the coin states $\{\ket{R}, \ket{L}\}$, representing heads and tails.
The evolution of the quantum walk for one step is governed by the unitary operator:

\begin{equation}\label{eq:unitary_qw_c}
    U = S \cdot (R_{c} \otimes I),
\end{equation}

\noindent
where $I$ is the identity operator in $\mathcal{H}_p$, and $R_{c}$ is a rotation in $\mathcal{H}_c$. 
In its generic matrix form, $R_c$ can be written as:

\begin{equation}\label{eq:generic_rotation_coin}
    R_c(\phi, \theta) = \begin{bmatrix} e^{i \phi} \cos(\theta) & e^{i \phi} \sin(\theta) \\ -e^{-i \phi} \sin(\theta) & e^{-i \phi} \cos(\theta) \end{bmatrix},
\end{equation}

\noindent
where $\phi$ denotes the phase angle of the rotation, and $\theta$ represents the actual rotation angle. 
The shift operator $S$ moves the walker one position to the right or left on the circle based on its coin state: if the coin is $\ket{R}$, the walker moves to position $x+1 \pmod{P}$, and if $\ket{L}$, to position $x-1 \pmod{P}$, for each $x \in \{0, \dots, P-1\}$.
Depending on the coin state, this operator shifts the walker either clockwise or counterclockwise. 
Since the walker is on a circle, position $P$ is identified with position $0$, creating a continuous loop of discrete steps.

%---------------------------------------------------------------------
\subsection{Hypercube-based quantum walks}\label{subsec:hypercube_quantum_walks}
%---------------------------------------------------------------------

\noindent
The representation of the hypercube model for \ac{qws} can be formulated based on the approach presented by Portugal~\cite{portugal_quantum_walks}.
In that case, the coin space, denoted as $\mathcal{H}_c = \mathcal{H}^{n}$, corresponds to the state of the quantum coin, which can be in a superposition of $n$ possible states. 
The walker space, denoted as $\mathcal{H}_p = \mathcal{H}^{2^{n}}$, represents the possible positions of the walker, with each position corresponding to a binary string of length $n$, leading to $2^{n}$ distinct positions. 
The total state of the system is the tensor product of these two spaces: $\mathcal{H} = \mathcal{H}_c \otimes \mathcal{H}_p$, where the coin state $\ket{a}$ satisfies $0 \le a < n$ and the walker position $\ket{v}$ ranges over all $2^n$ binary strings from $(00\dots00)_2$ to $(11\dots11)_2$.
The value of $a$ determines the position in the walker space where the next move occurs. 
Specifically, $a$ identifies which bit in the position vector $\ket{v}$ has to be flipped.
When the coin state is $\ket{a}$, the shift operator will flip the $a$-th bit of the walker's current position $\ket{v}$, thereby determining the walker's next position.
Regarding the coin operator, we can apply a generic $R_{c}$, as defined in Equation~\ref{eq:generic_rotation_coin}, to each coin qubit. 
However, for the hypercube topology, the optimal choice is theoretically the Grover coin~\cite{portugal_quantum_walks}, which can be compactly expressed as:

\begin{equation}\label{eq:grover_coin}
    G = \frac{2}{n}J_n - I_n,
\end{equation}

\noindent
where $n$ is the number of positions in the \ac{qw}, $I_n$ is the $n \times n$ identity matrix, and $J_n$ is the $n \times n$ all-ones matrix.
Now, let's examine the shift operator $S$, which determines how the walker moves across the hypercube based on the coin state. 
To formalize this, let $e_{a}$ represent a vector with a single $1$ in the $a$-th position and $0$ everywhere else. 
The shift operator $S$ is then defined as:

\begin{equation}\label{eq:shift_operator_h}
    S(\ket{a} \otimes \ket{v}) = \ket{a} \otimes \ket{v \oplus e_{a}},
\end{equation}

\noindent
where the system's state consists of two parts:

\begin{itemize}[label=\textbullet]
    \item $\ket{a}$: coin state, which determines the bit to modify. It specifies the position $a$ in the binary representation of the walker's position $\ket{v}$;
    \item $\ket{v}$: walker's current position, represented as a binary string corresponding to a node on the hypercube.
\end{itemize}

\noindent
The walker's position is updated by flipping the bit at position $a$. 
This is done using the binary XOR operation ($\oplus$) between the walker's position $v$ and a vector $e_{a}$. 
Consequently, after applying $S$, the coin state $\ket{a}$ remains unchanged, while the walker's position is updated to reflect the movement imposed by the coin.
Moreover, the final unitary operator $U$ governing the walk can be defined using Equation~\ref{eq:intro_unitary_qw}, where the coin operator $C$ is chosen as either the generic rotation coin (see Equation~\ref{eq:generic_rotation_coin}) or the Grover coin (see Equation~\ref{eq:grover_coin}).

%---------------------------------------------------------------------
\section{Protocol proposal}\label{sec:qkd_protocol}
%---------------------------------------------------------------------

\noindent
This section presents our one-way \ac{qkd} protocol using \ac{qws} on a $P$-dimensional hypercube within a prepare-and-measure framework. Building on Vlachou et al.~\cite{vlachou_qkd_quantum_walks}, we replace the circle topology with a hypercube, expanding the state space from $2P$ to $2^P$ basis states. This exponential growth enriches interference patterns, reduces predictability, and enhances resilience to eavesdropping and noise, improving security without requiring entanglement or quantum memory.

%---------------------------------------------------------------------
\subsection{Protocol description}\label{subsec:qkd_protocol_description}
%---------------------------------------------------------------------

\noindent
We begin by defining the shared public parameters $P$, $t$, $\phi$ and $\theta$, where:

\begin{itemize}[label=\textbullet]
    \item $P$: dimension of the position space of the quantum walk;
    \item $t$: number of steps performed in the \ac{qw} evolution;
    \item $\phi$, $\theta$: coin parameters.
\end{itemize}

\noindent
As previously defined in Equation~\ref{eq:unitary_qw_c}, we can express using $U$ the \ac{qw} operator, where $U$ is publicly known to all parties. 
Next, we introduce the $F$ operator, which acts on the coin space $\mathcal{H}_c$ before the \ac{qw} evolution, rather than as a post-processing operator.
This adjustment ensures that the coin state is prepared in a specific way before the \ac{qw} begins, rather than altering the coin state after the walk has been completed.
In our case, the operator $F$ can be either the identity ($I$), the Hadamard gate (denoted as $\tilde{X}$), or a cascade of a Hadamard gate followed by a phase gate ($Y = H \cdot S$), where $\tilde{X}$ and $Y$ are defined as follows:

\begin{equation}\label{eq:f_operator}
    \tilde{X} = H = \frac{1}{\sqrt{2}}\begin{bmatrix} 1 & 1 \\ 1 & -1\end{bmatrix}, \: Y = H \cdot S = \frac{1}{\sqrt{2}}\begin{bmatrix} 1 & 1 \\ i & -i\end{bmatrix}.
\end{equation}

\noindent
In general, the operator $F$ is optional, and when no transformation is applied, it is set to $F = I$.
For clarity, note that $\tilde{X}$ here refers to the Hadamard gate ($H$), not the classical $X$ gate, while $Y$ represents a combination of Hadamard and phase gates ($Y = H \cdot S$), used for coin state balancing.
This notation, adopted for consistency with Vlachou et al.~\cite{vlachou_qkd_quantum_walks}, allows us to explore different transformations on the coin state to fine-tune the \ac{qw} behavior.
Let $\ket{\psi_i}$ represent the protocol's state after the complete evolution, defined as:

\begin{equation}\label{eq:psi_i_one_way_qkd}
    \ket{\psi_i} = U^t \cdot (I \otimes F) \ket{i}, \: \text{for} \: \ket{i} \in \mathcal{H}_p \otimes \mathcal{H}_c.
\end{equation}

\noindent
Moreover, the orthonormal basis $\{\ket{\psi_i}\}$ is referred to as the $\mathcal{QW}$ basis, derived from the computational basis $\mathcal{Z}$ through the \ac{qw} evolution.
It is important to note that the unitary operator governing this basis change is the same unitary operator that describes the \ac{qw} process.
The protocol consists of $N$ iterations, each comprising the following steps:

\begin{enumerate}
    \item Alice (A) chooses a random bit $w_A \in  \{0,1\}$ and random $i_A \in \{0, \dots, 2^{P} - 1\}$. Then, depending on $w_A$ value:
    \begin{itemize}[label=\textbullet]
        \item If $w_A = 0$, Alice prepares and sends to Bob (B), over the public quantum channel, the $2^{P}$-dimensional state:\begin{equation}\label{eq:psi_i_prepare_0_c}
            \ket{\psi_i} = \ket{i_A}.
        \end{equation}
        \item If $w_A = 1$, Alice prepares and sends to Bob, over the public quantum channel, the $2^{P}$-dimensional state:\begin{equation}\label{eq:psi_i_prepare_1_c}
            \ket{\psi_i} = U^t \cdot (I \otimes F) \ket{i_A}.
        \end{equation}
    \end{itemize}
    \item Bob chooses a random bit $w_B \in \{0, 1\}$. Then, depending on $w_B$ value:
    \begin{itemize}[label=\textbullet]
        \item If $w_B = 0$, Bob measures the received state in the computational $\mathcal{Z}$ basis;
        \item If $w_B = 1$, Bob measures in the $\mathcal{QW}$ basis or, alternatively, he inverts the \acrshort{qw} by applying $(U^{t})^{-1} = U^{-t}$ and measures the resulting state in the $\mathcal{Z}$ basis.
    \end{itemize} 
    Let $j_B$ be the outcome in each case. 
    \item Alice and Bob reveal $w_A$ and $w_B$ via a classical authenticated channel. Then, based on their choices:
    \begin{itemize}[label=\textbullet]
        \item If $w_A = w_B$, $i_A$ and $j_B$ contribute to the raw key;
        \item If $w_A \neq w_B$, the iteration is discarded.
    \end{itemize}
\end{enumerate}

\noindent
After completing the process, Alice and Bob use a cut-and-choose method~\cite{yao_cut_and_choose} to detect eavesdropping by selecting a subset of iterations for parameter estimation and removing them from the raw key. 
This estimates the disturbances $Q_{z}$ and $Q_{w}$ in the $\mathcal{Z}$ and $\mathcal{QW}$ bases, which are ideally zero.
If disturbances remain below a defined threshold, they proceed with error correction and privacy amplification.
Finally, a scheme of the protocol is provided in Appendix~\ref{apn:qkd_protocol_schemes}.

%---------------------------------------------------------------------
\subsection{Security proof}\label{subsec:security_qkd}
%---------------------------------------------------------------------

\noindent
In this subsection, we prove the security of the proposed protocol by deriving an equivalent entanglement-based version, following standard \ac{qkd} techniques~\cite{quantum_crypto_without_bell,uncond_security_qkd}.
Establishing security for the entanglement-based protocol also validates the corresponding prepare-and-measure version~\cite{quantum_crypto_without_bell,entaglement_precond_secure_qkd,detecting_two_party_q_corr_qkd} and can extend to device-independent \ac{qkd} under suitable device assumptions~\cite{secrecy_pm_csh}.
The proof builds on Vlachou et al.~\cite{vlachou_qkd_quantum_walks} and can be adapted to hypercube-based \ac{qkd} with minor modifications.
For the entanglement-based protocol, each of the $N$ iterations modifies the initial steps as follows:

\begin{enumerate}
    \item Alice (A) prepares the entangled state:\begin{equation}\label{eq:psi_0_ent_based_protocol_h}
        \ket{\Psi_0} = \frac{1}{\sqrt{2^{P}}} \sum_{i=0}^{2^{P}-1} \ket{i, i}_{AB}.
    \end{equation}
    \noindent
    Then, she sends to Bob (B) the second portion of the prepared entangled state ($\ket{\Psi_0}_B$), while retaining the first portion ($\ket{\Psi_0}_A$) in her private lab;
    \item Alice and Bob independently choose two random bits, $w_A \in \{0, 1\}$ and $w_B \in \{0, 1\}$. If $w_A = 0$, Alice measures $\ket{\Psi_0}_A$ in the computational $\mathcal{Z}$ basis. Otherwise, she measures in the $\mathcal{QW}$ basis. Bob similarly measures $\ket{\Psi_0}_B$ according to $w_B$. Their measurement outcomes are recorded as $i_A$ for Alice and $j_B$ for Bob.
\end{enumerate}

\noindent
After these adjustments, the entanglement-based version proceeds as the prepare-and-measure counterpart, following the same basis reconciliation and subsequent steps.
Additionally, Appendix~\ref{apn:qkd_protocol_schemes} includes a depiction of the entanglement-based scheme.
Next, we demonstrate the security of the hypercube entanglement-based protocol by first making three assumptions:

\begin{itemize}[label=\textbullet]
    \item $A_1$: Alice and Bob only use iterations where $w_A = w_B = 0$ for their raw key;
    \item $A_2$: Eve is limited to collective attacks, where she independently attacks each protocol iteration but can perform a joint measurement of her ancilla at any future time;
    \item $A_3$: Eve prepares the states that Alice and Bob hold.
\end{itemize}

\noindent
Assumption $A_1$ simplifies computations and can be discarded later.
Alternatively, Alice and Bob can intentionally bias their selection of measurement bases to increase the probability that both choose $w_A = w_B = 0$, following a similar strategy used in the BB84 protocol.
Moreover, assumption $A_2$ can be removed later using a de-Finetti argument, yielding security in the asymptotic limit without degrading the key-rate~\cite{inf_theoretic_proof_qkd,postselection_tech_qc,symmetry_large_systems_ind_subsystem}.
It is worth noting that removing assumption $A_2$ is sufficient to establish the protocol's security.
Instead, assumption $A_3$ grants more power to Eve: if security is shown with $A_3$, it holds even when $A_3$ is removed.
Additionally, it is important to underline that this proof focuses exclusively on the asymptotic regime, where the key-rate expression is unaffected by finite-size effects, a standard assumption in theoretical \ac{qkd} security proofs~\cite{security_practical_qkd}, while finite-key analyses provide the necessary corrections for practical implementations~\cite{tomamichel_finite_key}.
Given $A_2$ and $A_3$, Alice, Bob, and Eve, after $N$ iterations, share a quantum state $\hat{\rho}_{ABE}^{\otimes N}$, where:

\[
    \hat{\rho}_{ABE} \in \mathcal{H}_A \otimes \mathcal{H}_B \otimes \mathcal{H}_E, \: \text{with} \: \mathcal{H}_A \equiv \mathcal{H}_B \equiv \mathcal{H}_p \otimes \mathcal{H}_c.
\]

\noindent
Note that Eve, as an all-powerful adversary, is not limited in the choice of her Hilbert space $\mathcal{H}_E$.
After information reconciliation and privacy amplification, Alice and Bob share a secret key of size $\ell(N)$.
Under the assumption of collective attacks ($A_2$), the Devetak-Winter key-rate expression~\cite{devetak_winter_key_rate} can be written explicitly as:

\begin{equation}\label{eq:devetak_winter_r}
    r = \lim_{N \to \infty} \frac{\ell(N)}{N} = S(A|E) - H(A|B).
\end{equation}

\noindent
Let $A_z$ and $A_w$ be random variables representing Alice's system when measured in the $\mathcal{Z}$ or $\mathcal{QW}$ basis, respectively, with $B_z$ and $B_w$ defined similarly for Bob.
Under assumption $A_1$, we are interested in:

\begin{equation}\label{eq:r_alice_system}
    r = S(A_z|E) - H(A_z|B_z).
\end{equation}

\noindent
Computing $H(A_z|B_z)$ is straightforward, given the probabilities:

\begin{equation}\label{eq:prob_shannon_entropy}
    p_{i,j}^z = \mathbb{P}(i_A = i, j_B = j \mid w_A = w_B = 0).
\end{equation}

\noindent
The challenge lies in bounding the Von-Neumann entropy $S(A_z|E)$.
To do this, we apply an uncertainty relation~\cite{uncertainty_principle} which states that, for any density matrix $\hat{\rho}_{ABE} \in \mathcal{H}_A \otimes \mathcal{H}_B \otimes \mathcal{H}_E$, if Alice and Bob perform measurements with POVMs:

\[
    \hat{M}_0 = \{\hat{M}_x^{(0)}\}_x \: \text{or} \: \hat{M}_1 = \{\hat{M}_x^{(1)}\}_x,
\]

\noindent
then:

\begin{equation}\label{eq:uncertainty_rel_entropy}
    S(A_0|E) + H(A_1|B) \geq \log \frac{1}{c},
\end{equation}

\noindent
where $c$ is given by:

\begin{equation}\label{eq:preliminary_c}
    c = \max_{x,y} \| \hat{M}_x^{(0)} \hat{M}_y^{(1)} \|_\infty^2,
\end{equation}

\noindent
considering $\|\cdot\|_\infty$ as the operator norm, with $A_i$ representing a random variable that describes Alice's system after applying the measurement $\hat{M}_i$.
Similarly, we can define $B_i$ for Bob's system.
Assuming measurements $\hat{M}_0$ are used for key distillation, we derive the following Devetak-Winter key-rate:

\[
    r = S(A_0|E) - H(A_0|B_0) \geq \log \frac{1}{c} - H(A_0|B_0) - H(A_1|B),
\]

\noindent
then, building on the principle that measurements can only increase entropy, we get:

\begin{equation}\label{eq:r_devetak_winter_final_2}
    r \geq \log \frac{1}{c} - H(A_0|B_0) - H(A_1|B_1).
\end{equation}

\noindent
In our scenario, we set the measurement operators as:

\[
    \hat{M}^{(0)}_x = \ket{x}\bra{x} \: \text{for} \: \mathcal{Z} \: \text{basis}, \:
    \hat{M}^{(1)}_x = \ket{\psi_x} \bra{\psi_x} \: \text{for} \: \mathcal{QW} \: \text{basis},
\]

\noindent
where, in the context of hypercube-based quantum walks, we consider $x \in \{0, 1, \dots, 2^{P} - 1\}$, and let $\ket{\psi_x}$ be defined as $\ket{\psi_x} = \sum_{i}\alpha_{x,i} \ket{i}$.
Then, it follows directly that:

\[
    \|\hat{M}^{(0)}_x \hat{M}^{(1)}_y\|_\infty^{2} = |\alpha_{x, y}|^2, \: \text{for all} \: x,y,
\]

\noindent
which leads to:

\begin{equation}\label{eq:final_c_security}
    c = \max_{x,y} |\alpha_{x, y}|^2,
\end{equation}

\noindent
where $c$ depends only on the \ac{qw} parameters, and remains unaffected by both channel noise and the protocol design.
It represents the maximum probability of any outcome $(x,y)$, so lower $c$ values yield a more uniform distribution (i.e., improved security), limiting an eavesdropper's ability to infer key information.
By choosing $F$, $t$, $\phi$, and $\theta$ appropriately, Alice and Bob can optimize both security and key rate.
Importantly, this proof applies to any regular topology, not just the hypercube, as it is independent of the specific graph structure.

%---------------------------------------------------------------------
\section{Qiskit models}\label{sec:qiskit_model}
%---------------------------------------------------------------------

\noindent
In this section, we first present the model underlying quantum walks and the subsequent \ac{qkd} protocols constructed on two distinct topologies: circle and hypercube.
Built upon the framework established by Douglas et al.~\cite{douglas_wang_eff_circuit_quantum_walks}, as also discussed in Section~\ref{sec:related_works}, this model structures the quantum walk using increment and decrement operations implemented with $X$ (i.e., \textsc{NOT}) gates and multi-controlled $X$ gates (i.e., \textsc{MCX}), which serve as the fundamental building blocks for both topologies.

%---------------------------------------------------------------------
\subsection{Qiskit implementation of quantum walks}\label{subsec:qiskit_impl_qws}
%---------------------------------------------------------------------

\noindent
In this subsection, we present the implementation of circle and hypercube-based \ac{qws}, with evolution defined as detailed in Subsections~\ref{subsec:circle_quantum_walks} and~\ref{subsec:hypercube_quantum_walks}.
In circle-based walks, the walker's position is encoded in $n = \lceil\log_2(2P)\rceil$ qubits, while a single coin qubit undergoes a unitary rotation parameterized by angles $\theta$ and $\phi$ (Equation~\ref{eq:generic_rotation_coin}).
The walker is initialized in the computational basis ($\mathcal{Z}$), and conditional shifts are applied: if the coin is $\ket{1}$, the walker moves right; if $\ket{0}$, left.
These shifts are implemented via multi-controlled $X$ gates, and an additional operator $F \in \{I, \tilde{X}, Y\}$ is applied to the coin before each step to balance the distribution.
The hypercube walk follows the same principles with a larger register: the walker's position is encoded by $P$ qubits, and $P$ coin qubits determine movement along each dimension.
Each step applies a coin operation (generic rotation or Grover coin) followed by dimension-dependent \textsc{MCX} shifts.
The $F$ operator and coin principles are applied as in the circle model, preserving modularity and extending naturally to higher dimensions.
As detailed in Appendix~\ref{apn:qiskit_implementation}, we provide the Qiskit implementations of the circle and hypercube-based quantum walk topologies. 
In both cases, the \ac{qws} are initialized with $X$ gates to set the states in the computational basis. 
An alternative is the Hadamard walk, where $H$ gates create superpositions~\cite{quantum_walks_review}, but this is unsuitable for our scenario: applying $H$ to $\ket{0}$ yields a superposition that remains unchanged after an $X$ gate, preventing proper initialization on a computational basis.  
Regarding measurement, the reversal ensures that the walker's readout matches the register, proceeding from the most significant bit (MSB) to the least significant bit (LSB).  
Finally, to approximate a randomized \ac{qw}, larger values of $t$ are required. 
Our simulations show that for $t \approx 10^{3}$ the distribution becomes nearly uniform across all walker states, but in the hypercube case this occurs only with a generic coin rotation rather than Grover's coin.

%---------------------------------------------------------------------
\subsection{Qiskit implementation of one-way QKD protocol}\label{subsec:qiskit_impl_qkd}
%---------------------------------------------------------------------

\begin{figure}[htbp]
     \centering
     \includegraphics[width=\columnwidth]{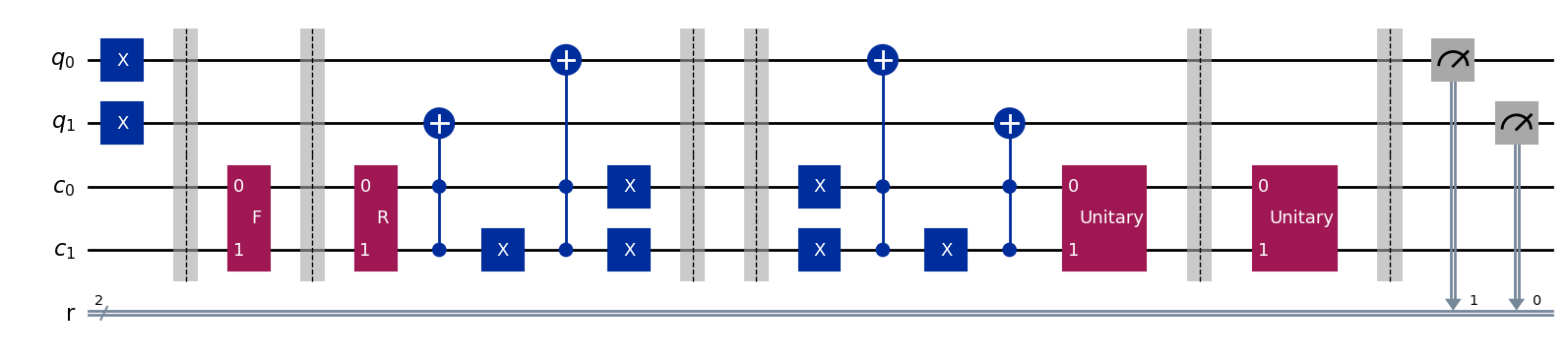}
     \caption{Example of a hypercube-based one-way \ac{qkd} protocol with $P = 2$, $t = 1$, and $N = 1$. When $w_A = 1$, Alice prepares a random initial state $\ket{i_A} = \ket{3} = \ket{11}_2$ from the $2^{P} = 2^2$ set and applies a \ac{qw} evolution on the hypercube. Bob then inverts the walk and measures in the computational basis to recover $\ket{i_A}$, as highlighted in the second part of the figure. If $w_A = 0$, Alice only prepares $\ket{i_A}$ without any transformation, and Bob directly measures it on a computational basis.}
     \label{fig:qkd_protocol_hypercube}
\end{figure}

\noindent
Building upon the quantum walk models and their implementation introduced in the previous subsection, we present the Qiskit-based implementation of our one-way \ac{qkd} protocol on both circle and hypercube topologies.
These realizations reuse the circuits and operators defined earlier, showing how the same \ac{qw} formalism extends naturally to secure communication.
Following the prepare-and-measure scheme of Section~\ref{sec:qkd_protocol}, Alice encodes information by applying a sequence of walk steps to a known initial state (for $w_A = 1$).
Bob then applies Qiskit's inverse function, which computes the exact inverse of a circuit step by step~\cite{qiskit_inverse}, effectively undoing Alice's operations and allowing him to recover the walker's logical position via measurement in the computational basis.
This mechanism avoids the need for quantum memory and ensures precise decoding of the encoded state.
When both choose $w_A = w_B = 0$, no walk is performed: Alice prepares $\ket{i_A}$, and Bob measures directly. The case requiring full quantum walk dynamics arises only when $w_A = w_B = 1$.
Figure~\ref{fig:qkd_protocol_hypercube} illustrates the complete hypercube implementation, highlighting the modularity of the approach, which scales efficiently by adjusting the number of qubits and walk layers while preserving the protocol's logic and security guarantees.

%---------------------------------------------------------------------
\section{Security Evaluation}\label{sec:security_evaluation}
%---------------------------------------------------------------------

\noindent
In this section, we assess the security of our proposed \ac{qkd} protocol.
As discussed in Subsection~\ref{subsec:security_qkd}, the measurement overlap $c$ (Equation~\ref{eq:final_c_security}) is fully determined by the quantum walk parameters selected by Alice and Bob.
Thus, they should aim to select a \ac{qw} that minimizes this value, ensuring that after $t$ steps of evolution, the probability of finding the walker at any specific position is low (ideally resulting in a uniform distribution).
Notably, as $t \to \infty$, the values $|\alpha_{x,y}|$ do not settle into a steady state.
This is why \ac{qws} on regular structures are typically analyzed using the time-averaged distribution~\cite{quantum_walks_review}.
In our \ac{qkd} protocol, we do not focus on large $t$, but rather on finding an optimal $t$ that is not too high, as increasing $t$ makes more time-consuming Alice's state preparation and Bob's reversal.
A larger $t$ does not necessarily make it more difficult for Eve to distinguish states but helps achieve a more uniform probability distribution, reducing state predictability.
Since $t$ affects noise tolerance, we aim to determine the value that maximizes noise resistance for a given walk configuration.
In practice, ``optimal'' would likely differ, since Alice and Bob would need to account for device imperfections.
All simulations (both in this section and the following ones) were conducted on a \ac{vm} equipped with a 12-core 2.0\,GHz Intel Core (Haswell) CPU and 31\,GiB RAM.
We analyze different walk parameters to determine the minimum value $c$ when $F = I$.
To achieve this, we developed multiple scripts to simulate the \ac{qw} using the Qiskit models defined in Section~\ref{sec:qiskit_model}.
We run the simulations for \ac{qw} steps $t$ ranging from $1$ to $5 \times 10^{4}$ to identify the optimal $t$ that minimizes $c$.
For this evaluation, we used the generic coin rotation operator (consider Equation~\ref{eq:generic_rotation_coin}) with $\phi = 0$ and $\theta = \pi/4$.
In the following subsections, we begin by reproducing and validating the state-of-the-art results presented by Vlachou et al.~\cite{vlachou_qkd_quantum_walks}, using a Qiskit-based implementation that significantly differs from the original model adopted by the authors.
Subsequently, we enhance the protocol's security performance by introducing quantum walks on a hypercube topology.

%---------------------------------------------------------------------
\subsection{Verification of state-of-the-art results}\label{subsec:sota_verification}
%---------------------------------------------------------------------

\begin{figure}[htbp]
    \centering
    \includegraphics[width=0.9\columnwidth]{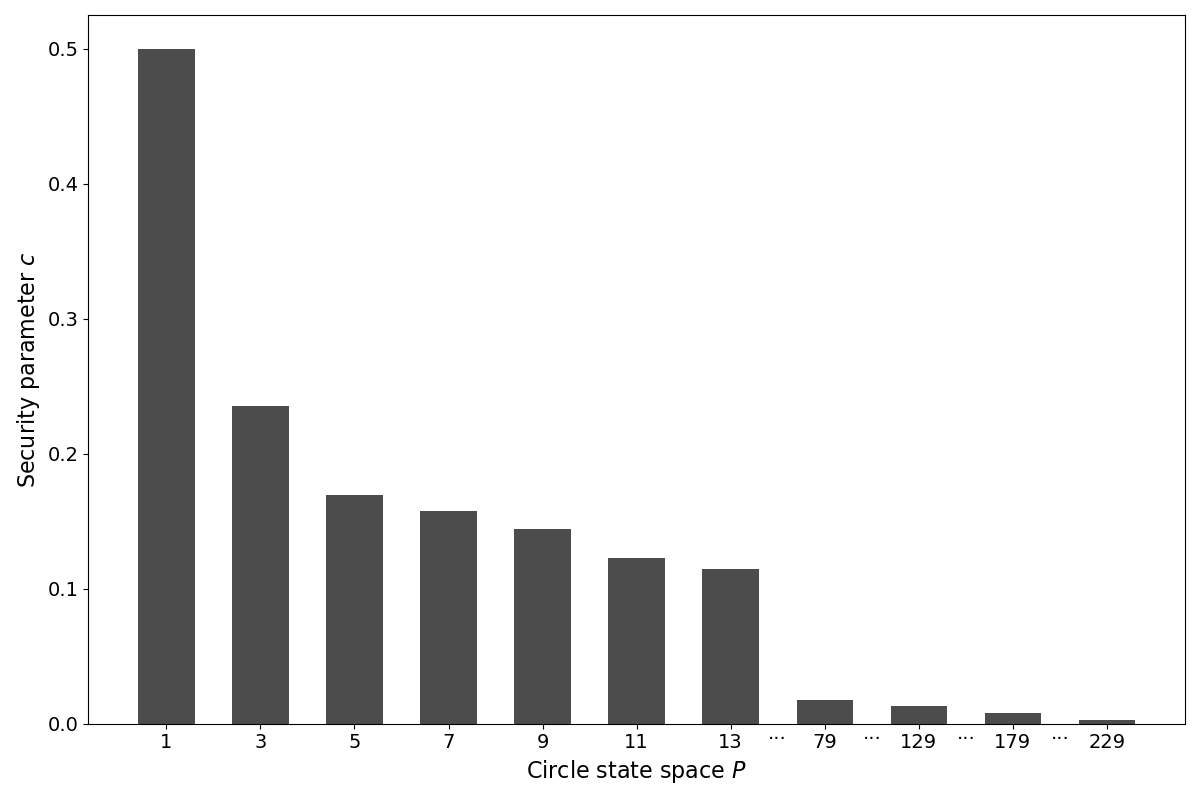}
    \caption{Minimal measurement overlap $c$ obtained for each position space dimension $P$ in the circle topology, using $\phi = 0$, $\theta = \pi/4$, and $F = I$. It is important to note that a smaller value of $c$ is more advantageous for Alice and Bob. Additionally, $P$ represents the dimension of the position space. This serves as our state-of-the-art, from which we begin our analysis to improve performance.}
    \label{fig:c_analysis_circle_all}
\end{figure}

\noindent
In this subsection, we first replicate the state-of-the-art results reported by Vlachou et al.~\cite{vlachou_qkd_quantum_walks}, comparing their $c$ and optimal $t$ values with those obtained from our hypercube-based \ac{qws} to establish a reference for our analysis.
The results of our initial evaluation are presented in Figure~\ref{fig:c_analysis_circle_all}.

\noindent
It is important to highlight that we replicate the results from Vlachou et al.~\cite{vlachou_qkd_quantum_walks} using a Qiskit-based model, which represents an entirely different approach compared to the simulations performed before this work.
However, the results are consistent, given the differences in the two methods.
As can be observed from Figure~\ref{fig:c_analysis_circle_all}, we considered odd values of $P$ ranging from $1$ to $229$ in order to align with the original analysis presented by Vlachou et al.~\cite{vlachou_qkd_quantum_walks}, since only odd values of $P$ are meaningful in the circle topology~\cite{quantum_walks_graphs}.
Using even values of $P$ would restrict the support of the probability amplitudes to either even or odd numbered nodes, increasing the overall value of $|\alpha_{x,y}|$.
In addition, for $P = 1$, we obtain $c = 0.5$, which matches the expected result for a classical BB84 protocol (i.e., when $P = 1$, our model collapses into the BB84).
For all other values of $P$, as $P$ increases, we observe smaller values of $c$ with a reasonable $t$.

%---------------------------------------------------------------------
\subsection{Hypercube-based quantum walks results}\label{subsec:hypercube_qws_results}
%---------------------------------------------------------------------

\begin{figure}[htbp]
    \centering
    \includegraphics[width=0.9\columnwidth]{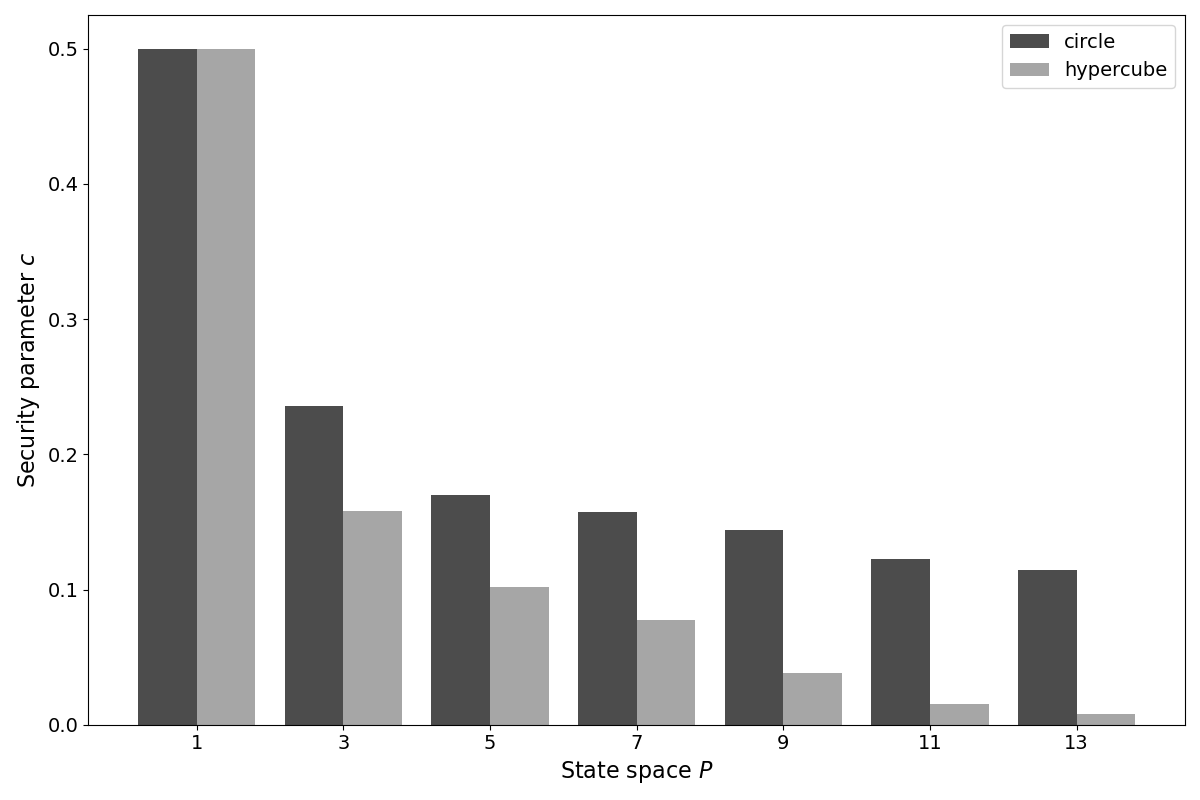}
    \caption{Comparison plot between hypercube and circle-based quantum walks using the same parameters. The minimal value of $c$ for each position space dimension $P$ is shown for both topologies, with $\phi = 0$, $\theta = \pi/4$, and $F = I$. It's important to note that Alice and Bob are better off with a smaller $c$, and $P$ denotes the position space dimension.}
    \label{fig:c_analysis_comparison}
\end{figure}

\noindent
In this subsection, we analyze the security of our protocol when the underlying \ac{qws} are performed on the hypercube topology.
In this configuration, the state space grows exponentially with $P$, reaching a dimension of $2^{P}$, in contrast to the linear growth $2P$ in the circle.
As a result, the probability amplitude naturally spreads across many possible outcomes, reducing the likelihood that any single outcome becomes dominant.
This wider distribution is expected to lower the value of $c$, thereby enhancing the protocol's efficiency by making the quantum states more difficult to predict, since security itself already follows from privacy amplification.
To allow a more precise comparison, we used the same parameters as in the circle-based simulations when evaluating the hypercube-based case.
However, due to the limitations of the AerSimulator in the Qiskit environment~\cite{qiskit_aer}, we are unable to simulate hypercube-based \ac{qws} for $P > 13$.
Therefore, we will compare the results for $P \leq 13$, as shown in Figure~\ref{fig:c_analysis_comparison}.

\noindent
As specified earlier, our \ac{qkd} protocol aims to produce a uniform distribution of states, reducing predictability for an eavesdropper.
The Grover coin, although theoretically optimal for mixing speed~\cite{portugal_quantum_walks}, introduces a bias tied to the public initialization state~\cite{quantum_walks_grover_coin} in our one-way setting, making outcomes predictable with high probability. Referring to Equation~\ref{eq:final_c_security}, with $P = 1$ we expect $c = 0.5$, while larger $P$ should lower $c$ given sufficient iterations.
Yet, with Grover coin on hypercubes, $c$ remains fixed at $0.5$ even as $P$ grows, effectively collapsing the protocol to BB84.
For this reason, we exclude Grover coin-based \ac{qws} from further study, noting this is an experimental conclusion rather than a formal proof.
In contrast, hypercube-based walks with a generic rotation coin significantly outperform circle-based ones, yielding much lower $c$ values with a reasonable number of steps (i.e., $t \approx 10^{3}$).
This advantage comes from the exponentially larger and structured state space, which spreads probability more uniformly and limits distribution on specific outcomes.
However, this comes at the cost of increased implementation complexity, as the hypercube-based \ac{qws} requires a higher number of qubits. Therefore, a trade-off must be considered between implementation feasibility and security, which is notably enhanced through the hypercube topology.
To conclude, Table~\ref{tab:c_comparison_circle_hypercube} compares the two topologies, highlighting their differences and improvements.

\begin{table}[htbp]
    \centering
    \setlength{\tabcolsep}{5pt}
    \caption{Comparison of the minimal values of $c$ and optimal $t$ for circle-based and hypercube-based quantum walks \ac{qkd} protocol with $\phi = 0$, $\theta = \pi/4$, and $F = I$.}
    \begin{tabular}{c|cc|cc|}
        & \multicolumn{2}{|c|}{\textbf{circle}} & \multicolumn{2}{|c|}{\textbf{hypercube}} \\
        \cline{2-5}
        $P$ & $c$ & optimal $t$ & $c$ & optimal $t$ \\
        \hline
        1  & 0.5 & 1 & 0.5 & 1 \\
        3  & 0.236 & 1084 & 0.158 & 1492 \\
        5  & 0.170 & 1196 & 0.102 & 1162 \\
        7  & 0.157 & 1202 & 0.077 & 965 \\
        9  & 0.144 & 1032 & 0.038 & 1085 \\
        11 & 0.123 & 1594 & 0.0154 & 1283 \\
        13 & 0.114 & 932  & 0.008 & 1378 \\
    \end{tabular}
    \label{tab:c_comparison_circle_hypercube}
\end{table}

%---------------------------------------------------------------------
\section{Noise Resistance Analysis}\label{sec:noise_resistance_analysis}
%---------------------------------------------------------------------

\noindent
In this section, we analyze the robustness of our \ac{qkd} protocol under noise.
Using the parameters from the security analysis, we determine the optimal \ac{qw} configuration for a given $P$ and the corresponding $c$, then compute the secret key rate $r$ and identify noise levels where $r > 0$.
A physical implementation is not yet possible; otherwise, one could measure $p_{i,j}^{z}$ and $p_{i,j}^{w}$ (Equation~\ref{eq:prob_shannon_entropy}) to calculate $H(A_z|B_z)$ and $H(A_w|B_w)$ for Equation~\ref{eq:r_alice_system}.
Since such an implementation does not yet exist, we instead use noise models to estimate the \ac{qer}, a standard approach in the literature~\cite{qer_estimate_qkd,vlachou_qkd_quantum_walks}.

%---------------------------------------------------------------------
\subsection{Non-optimal protocol robustness}\label{subsec:protocol_robustness}
%---------------------------------------------------------------------

\begin{figure}[htbp]
    \centering
    \subfloat[Depolarizing noise\label{fig:qer_comparison_dn}]{%
    \includegraphics[width=0.48\columnwidth]{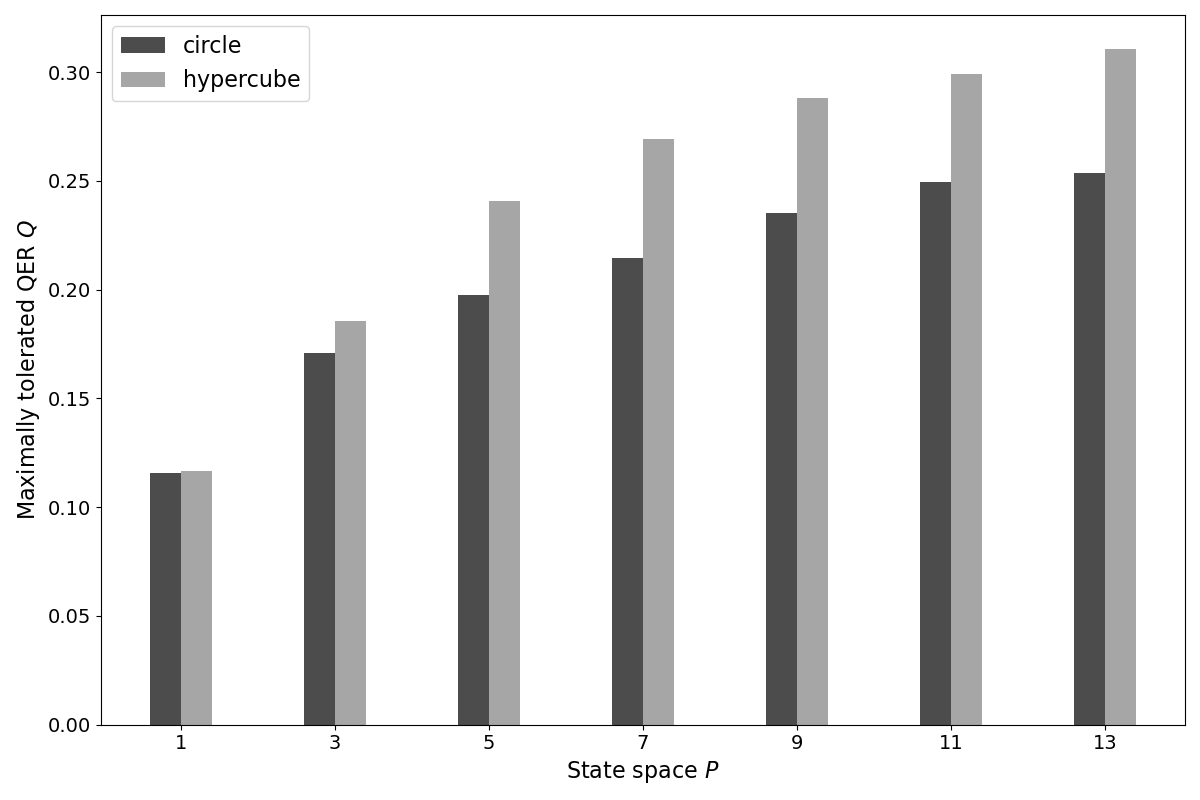}%
    }\hfill
    \subfloat[Amplitude-phase damping noise\label{fig:qer_comparison_dmp}]{%
    \includegraphics[width=0.48\columnwidth]{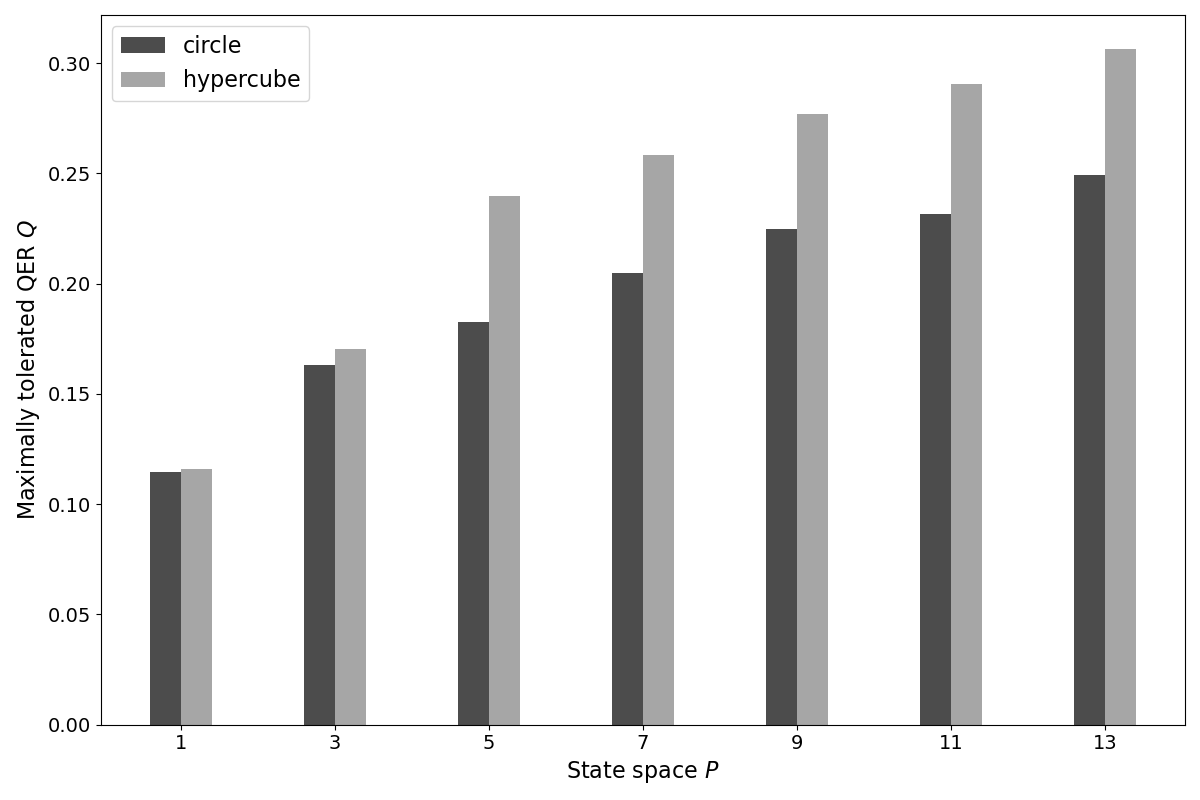}%
    }
    \caption{Comparison of the maximally tolerated \ac{qer} for circle and hypercube-based \ac{qkd} protocols under depolarizing and amplitude-phase damping noise, using $F = I$, $\phi = 0$, and $\theta = \pi/4$. For $P = 1$, the BB84 limit is recovered, while at $P = 13$, the hypercube-based protocol achieves $Q \approx 0.311$ under depolarizing noise and $Q \approx 0.306$ under amplitude-phase damping, showing improved noise tolerance.}
    \label{fig:qer_comparison_dep_amp}
\end{figure}

\noindent
This subsection describes the noise models used to evaluate the maximum tolerable 
\ac{qer} for a positive key rate (i.e., $r > 0$) and to assess protocol robustness.
For replicability, we employ Qiskit's built-in single-qubit noise models: 
depolarizing error~\cite{qiskit_depolarizing_error} and combined amplitude-phase 
damping error~\cite{noise_damping,qiskit_amplitude_damping,qkd_noisy_channels,quantum_error_correction_qubit}, 
which simulate realistic quantum channel noise~\cite{robustness_channel_noise}.
The depolarizing channel models random gate errors by randomly replacing the 
qubit state with the maximally mixed state with probability $\lambda$, 
effectively simulating imperfect gate operations~\cite{qiskit_depolarizing_error}.
Amplitude damping describes energy loss such as photon loss or qubit 
relaxation~\cite{excitation_damping_quantum_channels,amplitude_damping_codes}, 
while phase damping captures dephasing, reducing coherence without changing 
populations~\cite{qkd_noisy_channels,quantum_error_correction_qubit,qiskit_phase_damping}.
The combined amplitude-phase damping channel composes the two, capturing both 
energy loss and dephasing within a single unified noise model.
Errors are applied uniformly to all qubits via 
add\_all\_qubit\_quantum\_error~\cite{qiskit_noise_models}, affecting both walker 
and coin registers. While less precise than custom multi-qubit 
models~\cite{vlachou_qkd_quantum_walks}, this approach ensures consistency 
across qubits and straightforward comparability of results.
The maximally tolerated \ac{qer}, measuring discrepancies between Alice and Bob when using the same basis, is defined as:

\begin{equation}\label{eq:qer_estimate_noise}
    Q = \sum_{a \neq b} p^{z}_{a,b} = \mathbb{P}(A_z = a, B_z = b \mid w_A = w_B = 0),
\end{equation}

\noindent
which represents the probability that Alice and Bob obtain different outcomes when both measure in the $\mathcal{Z}$ basis~\cite{vlachou_qkd_quantum_walks}.
Since \ac{qer} accounts for noise and potential eavesdropping, it is a key security metric in our \ac{qkd} protocol.
In fact, a high \ac{qer} indicates excessive noise or an adversary's interference.
As a consequence, if it surpasses a critical threshold (i.e., $11\%$ in BB84), error correction and privacy amplification become ineffective, making secure key extraction impossible.
In our simulation, we use a depolarizing channel with a parameter $\lambda$ set to match the maximum noise tolerance observed in the BB84 protocol (i.e., when $P = 1$ and $t = 1$), corresponding to $Q_{\text{dep}} \approx 0.116$ (Table~\ref{tab:qer_comparison}), consistent with BB84's $11\%$ \ac{qer} bound.
For higher $P$ values, which correspond to a \ac{qw}-based evolution, the same $\lambda$ value is used to simulate the \ac{qkd} protocol, ensuring consistent evaluation across different topologies and parameter settings.
The same approach is applied to the amplitude-phase damping noise model simulations, where the amplitude and phase damping parameters were selected based on the same reasoning.
Let's examine the results obtained for the hypercube-based and circle-based \ac{qkd} protocols under depolarizing and amplitude-phase damping noise.
The outcomes of the simulations under the two noise models are shown in Figure~\ref{fig:qer_comparison_dep_amp}.

\noindent
The results of Vlachou et al.~\cite{vlachou_qkd_quantum_walks} differ slightly, as they used a customized generalized Pauli noise model, better suited to their setup.
For our simulations, the number of iterations was fixed at $N = 10^{4}$ to ensure reliable estimates of the \ac{qer}, given the need for sufficient cases where $w_A = w_B = 0$.
As shown in Figure~\ref{fig:qer_comparison_dep_amp}, the hypercube-based \ac{qkd} protocol shows higher noise tolerance than the circle-based one.
In the circular case, each node connects to only two neighbors ($2P$ states), keeping the walk localized longer and making it more vulnerable to accumulated errors~\cite{localization_quantum_walks}.
In contrast, the hypercube connects each node to $P$ neighbors ($2^P$ states), allowing exponential spreading that distributes noise more evenly and reduces its impact~\cite{coherent_loc_coined_quantum_walks}.
Under depolarizing noise, the circle's confinement leads to correlated errors~\cite{qkd_intensity_fluctuations}, whereas the hypercube spreads states across a larger Hilbert space, mitigating noise per state.
A similar effect occurs with amplitude-phase damping, where decoherence distributes over more dimensions~\cite{qkd_coh_states_amp_dmp}.
Overall, the hypercube structure enhances resilience by enabling efficient error distribution, faster mixing, and reduced decoherence, thus supporting secure key generation at higher noise levels.
Finally, detailed results are reported in Table~\ref{tab:qer_comparison}.

\begin{table}[htbp]
    \centering
    \setlength{\tabcolsep}{5pt}
    \caption{Comparison of the maximally tolerated \ac{qer} results for circle-based and hypercube-based \ac{qkd} protocols under the two noise types: depolarizing noise ($Q_{\text{dep}}$) and amplitude-phase damping noise ($Q_{\text{dmp}}$), with parameters fixed as $F = I$, $\phi = 0$, and $\theta = \pi / 4$.}
    \begin{tabular}{c|ccc|ccc|}
        & \multicolumn{3}{|c|}{\textbf{circle}} & \multicolumn{3}{|c|}{\textbf{hypercube}}\\
        \cline{2-7}
        $P$ & optimal $t$ & $Q_{\text{dep}}$ & $Q_{\text{dmp}}$ & optimal $t$ & $Q_{\text{dep}}$ & $Q_{\text{dmp}}$ \\
        \hline
        1  & 1 & 0.116 & 0.115 & 1 & 0.117 & 0.116 \\
        3 & 1084 & 0.171 & 0.163 & 1492 & 0.185 & 0.170 \\
        5 & 1196 & 0.198 & 0.183 & 1162 & 0.241 & 0.240 \\
        7 & 1202 & 0.215 & 0.205 & 965 & 0.269 & 0.258 \\
        9 & 1032 & 0.235 & 0.225 & 1085 & 0.288 & 0.277 \\
        11 & 1594 & 0.250 & 0.231 & 1283 & 0.299 & 0.290 \\
        13 & 932  & 0.253 & 0.249 & 1378 & 0.311 & 0.306 \\
    \end{tabular}
    \label{tab:qer_comparison}
\end{table}

%---------------------------------------------------------------------
\subsection{Parameters optimization}\label{sub:params_optimization}
%---------------------------------------------------------------------

\noindent
This subsection analyzes the selection of optimal parameters for our \ac{qkd} protocol to maximize security (Section~\ref{sec:security_evaluation}) and noise resistance (Section~\ref{sec:noise_resistance_analysis}).
We minimize $c$ and maximize the tolerated \ac{qer} ($Q$) by adjusting the operator $F$ (from $I$ to $\tilde{X}$ or $Y$) to place the coin qubit(s) in superposition, and varying $\theta, \phi = \{k\pi / 10 \mid k = 0, \dots, 10\}$ for the coin rotation. Optimal parameters are identified by first minimizing $c$ for maximal security, then applying them in simulations of the circle or hypercube-based \ac{qkd} protocol to evaluate the maximally tolerated \ac{qer} under the noise models of Section~\ref{sec:noise_resistance_analysis}.
The optimal parameter combinations for both topologies are reported in Table~\ref{tab:opt_parameters_combination}.

\begin{table}[htbp]
    \centering
    \setlength{\tabcolsep}{5pt}
    \caption{Optimal parameter settings $F$, $\phi$, $\theta$ minimizing $c$ and corresponding optimal number of steps $t$ for both circle (top) and hypercube (bottom) configurations across different $P \neq 1$ values.}
    \begin{tabular}{c|ccccc}
        \multicolumn{6}{c}{\textbf{circle}} \\
        \hline
        $P$ & $F$ & $\phi$ & $\theta$ & $c$ & optimal $t$ \\
        \hline
        3  & $Y$        & 0        & $0.3\pi$ & 0.167 & 1322 \\
        5  & $Y$        & 0        & $0.5\pi$ & 0.106 & 1466 \\
        7  & $Y$        & $0.5\pi$ & $0.7\pi$ & 0.083 & 1638 \\
        9  & $\tilde{X}$& $0.5\pi$ & $0.8\pi$ & 0.068 & 1296 \\
        11 & $Y$        & $0.5\pi$ & $0.3\pi$ & 0.055 & 1374 \\
        13 & $Y$        & $0.5\pi$ & $0.9\pi$ & 0.046 & 1503 \\
        \hline
        \multicolumn{6}{c}{\textbf{hypercube}} \\
        \hline
        3  & $Y$        & 0        & $0.5\pi$ & 0.109 & 1514 \\
        5  & $\tilde{X}$& $0.2\pi$ & $0.8\pi$ & 0.080 & 1398 \\
        7  & $Y$        & $0.5\pi$ & $0.4\pi$ & 0.045 & 1611 \\
        9  & $Y$        & $0.4\pi$ & $0.6\pi$ & 0.017 & 1235 \\
        11 & $Y$        & $0.2\pi$ & $0.8\pi$ & 0.009 & 1750 \\
        13 & $Y$        & $0.5\pi$ & $0.3\pi$ & 0.004 & 1198 \\
    \end{tabular}
    \label{tab:opt_parameters_combination}
\end{table}

\begin{figure}[htbp]
    \centering
    \subfloat[Depolarizing noise\label{fig:qer_comparison_dep_final}]{%
    \includegraphics[width=0.48\columnwidth]{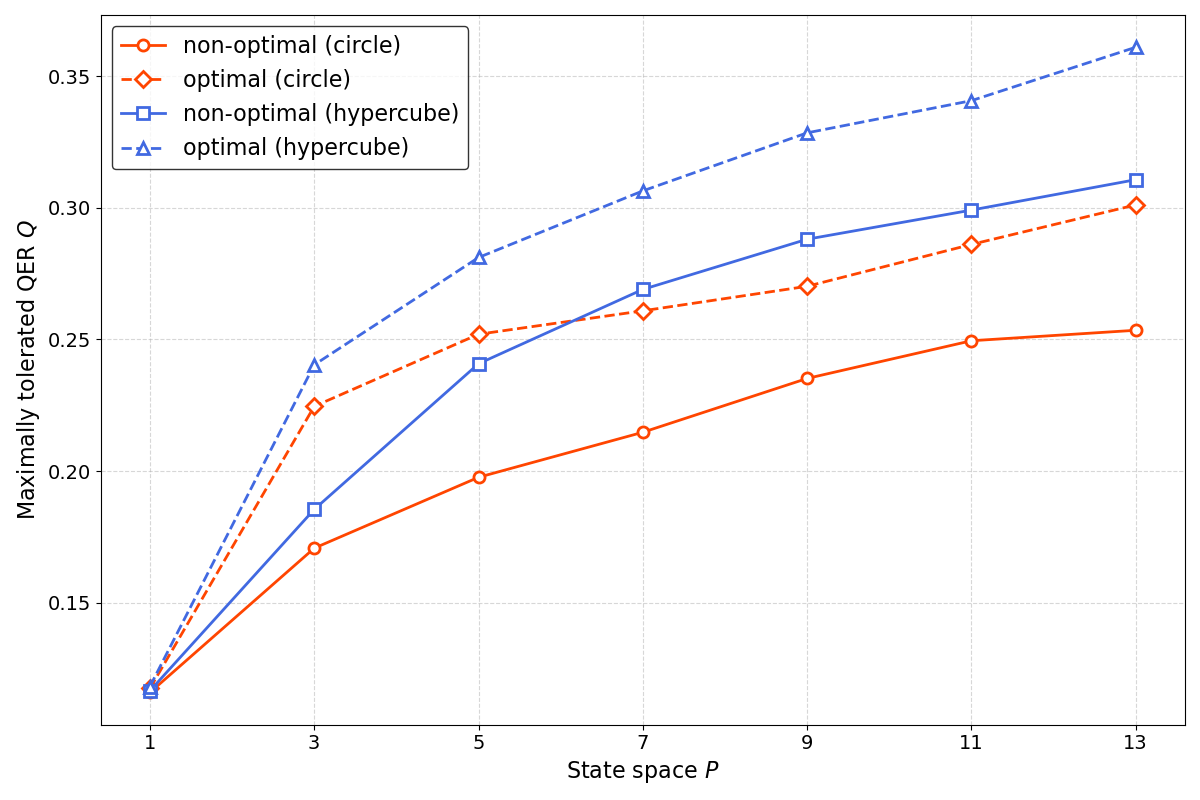}%
    }\hfill
    \subfloat[Amplitude-phase damping noise\label{fig:qer_comparison_dmp_final}]{%
    \includegraphics[width=0.48\columnwidth]{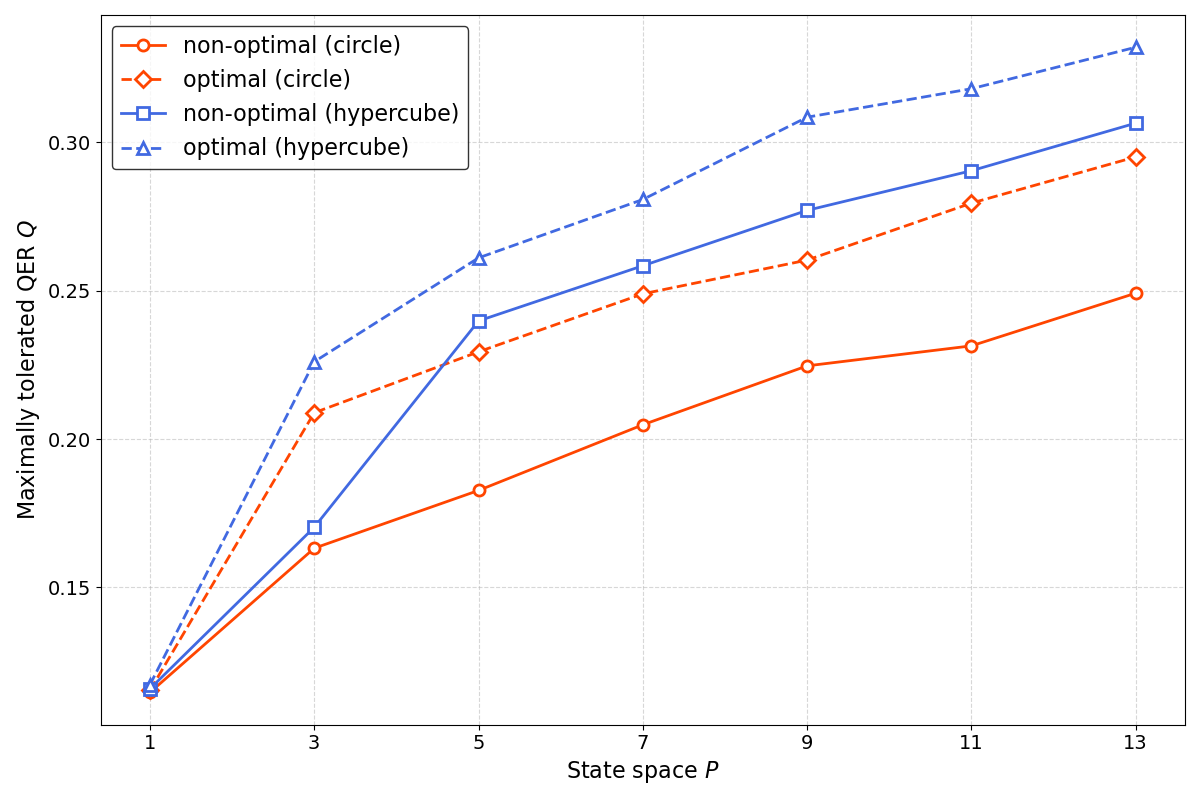}%
    }
    \caption{Comparison of the maximum tolerated \ac{qer} for circle-based and hypercube-based \ac{qkd} under different noise types, using optimal parameters from Table~\ref{tab:opt_parameters_combination}. The hypercube-based protocol shows greater error tolerance: up to $Q \approx 0.361$ (depolarizing) and $Q \approx 0.332$ (amplitude-phase damping).}
    \label{fig:qer_comparison_final}
\end{figure}

\noindent
From Table~\ref{tab:opt_parameters_combination}, optimal performance consistently requires $F \neq I$.
Choosing $F = \tilde{X}$ or $F = Y$ produces nearly uniform output distributions, crucial for security, whereas $F = I$ generally degrades protocol performance.
In Figure~\ref{fig:qer_comparison_final}, we compare the maximum tolerated \ac{qer} for circle-based and hypercube-based \ac{qkd} protocols under depolarizing and amplitude-damping noise models, utilizing the optimal parameters listed in Table~\ref{tab:opt_parameters_combination}. 

\noindent
As shown in Figure~\ref{fig:qer_comparison_final}, with an increase in $P$, both topologies exhibit an increase in \ac{qer}, with the hypercube consistently tolerating higher \ac{qer} values, reflecting its better noise resilience.
This advantage is most evident at smaller $P$, where the system transitions from a BB84 protocol (i.e., $P = 1$) to a multi-qubit \ac{qkd} protocol.
The hypercube's higher dimensionality offers better noise tolerance, keeping \ac{qer} more stable at smaller $P$ for secure \ac{qkd} protocols.
Moreover, parameter optimization improves performance, with optimal cases outperforming non-optimal ones.
While the hypercube continues to perform better at larger $P$, the difference decreases as system complexity grows.
These trends are consistent across various values of $\phi$ and $\theta$, highlighting the importance of topology and optimization in \ac{qer}, particularly during the transition to multi-qubit \ac{qkd} protocols.
In conclusion, Table~\ref{tab:opt_results_final} presents the optimal \ac{qer} results for each $P$ along with the corresponding parameters for our \ac{qkd} protocol under both noise conditions.

\begin{table}[htbp]
    \centering
    \setlength{\tabcolsep}{5pt}
    \caption{Optimal values for $c$, $Q_\text{dep}$, and $Q_\text{dmp}$ in circle and hypercube configurations shown in Table~\ref{tab:opt_parameters_combination}, with $Q_\text{dep}$ and $Q_\text{dmp}$ representing the maximally tolerated \ac{qer} for depolarizing and amplitude-phase damping noise, respectively.}
    \begin{tabular}{c|ccc|ccc}
        & \multicolumn{3}{|c|}{\textbf{circle}} & \multicolumn{3}{|c}{\textbf{hypercube}} \\
        \cline{2-7}
        $P$ & $c$ & $Q_\text{dep}$ & $Q_\text{dmp}$ & $c$ & $Q_\text{dep}$ & $Q_\text{dmp}$ \\
        \hline
        3 & 0.167 & 0.225 & 0.209 & 0.109 & 0.240 & 0.226 \\
        5 & 0.106 & 0.252 & 0.229 & 0.080 & 0.281 & 0.261 \\
        7 & 0.083 & 0.261 & 0.249 & 0.045 & 0.306 & 0.281 \\
        9 & 0.068 & 0.270 & 0.260 & 0.017 & 0.329 & 0.308 \\
        11 & 0.055 & 0.286 & 0.280 & 0.009 & 0.341 & 0.318 \\
        13 & 0.046 & 0.301 & 0.295 & 0.004 & 0.361 & 0.332 \\
    \end{tabular}
    \label{tab:opt_results_final}
\end{table}

%---------------------------------------------------------------------
\section{Discussion and Future Work}\label{sec:discussion}
%---------------------------------------------------------------------

\noindent
In this section, we delve into the limitations of this research and propose directions for future work to build upon our results.
We specifically focus on our proposed hypercube-based quantum walks \ac{qkd} protocol, discussing its current constraints, potential vulnerabilities to practical attacks, and then exploring directions for further improvement and theoretical analysis.

\noindent
\textbf{Limitations and practical attacks}. While our protocol proves more secure and noise-resistant than state-of-the-art schemes, several limitations still need to be addressed.
First, due to AerSimulator constraints, we could only simulate up to $P \leq 13$; larger $P$ may further improve performance, though saturation in \ac{qer} is possible.
The hypercube protocol also requires more qubits, raising computational costs compared to the circle-based version.
Moreover, our analysis relied on Qiskit's single-qubit noise models for reproducibility, but a tailored multi-qubit model would yield a more realistic evaluation.
Increasing iterations ($N \gg 10^{4}$) would improve \ac{qer} estimates, though our hardware limited this, as would access to a true quantum RNG in place of the pseudo-randomness used for $w_A, w_B$, and $\ket{i_A}$.
From a security perspective, the protocol does not depend on quantum memory: in both the prepare-and-measure and entanglement-based versions, Eve requires a stable memory to attack, while Alice and Bob do not.
Against practical threats such as photon-number-splitting~\cite{pns_attack}, decoy states~\cite{decoy_states} remain effective, and our one-way communication model supports this countermeasure.
However, authentication of the classical reconciliation channel is still essential to prevent man-in-the-middle attacks.
Other known attacks, including Trojan Horse~\cite{trojan_horse}, detector blinding, and time-shift~\cite{detector_blinding_time_shifts}, can be mitigated with standard countermeasures~\cite{trojan_horse_solution,detector_blinding_solution,time_shifts_solution}. Nonetheless, since our protocol involves more quantum states than traditional ones, its practical implementation may introduce new challenges that merit further study.  We also note that comparing circle and hypercube at the same $P$ does not separate the effect of topology from the effect of dimension, since the two spaces have different sizes; a matched-dimension study (circle at $P$, hypercube at $\log_2(2P)$) is left for future work.

\noindent
\textbf{Future work}. Although our Qiskit model is functionally correct, more efficient implementations of \ac{qws} could be explored, as our choice of \textsc{MCX} gates and the coined model was guided mainly by feasibility; alternative quantum walk models~\cite{quantum_walks_review} might yield better performance.
While we avoided the Grover coin due to its unsuitability in one-way \ac{qkd}, investigating other coin operators beyond a simple rotation could be worthwhile.
Our security proof (Subsection~\ref{subsec:security_qkd}) could also be strengthened, for example by deriving analytical expressions for optimal \ac{qw} parameters or for $c$ in Equation~\ref{eq:final_c_security}.
Another open question is the asymptotic noise tolerance as the walk dimension grows (i.e., $P \to \infty$, possibly $t \to \infty$): prior work on high-dimensional \ac{qkd} (without walks) showed tolerance up to $50\%$ disturbance~\cite{qkd_infinite_states}, raising the question of whether our protocol approaches similar thresholds.
Our analysis is limited to the sifting phase; information reconciliation could use classical methods such as Winnow~\cite{winnow_protocol}, LDPC codes~\cite{mondin_ldpc_codes}, BCH codes~\cite{traisilanun_bch_codes} or Cascade~\cite{cascade_protocol}, followed by privacy amplification via universal hash functions~\cite{bennett_privacy_amplification} or Toeplitz-based constructions~\cite{krawczy_lfsr_hashing_auth} to remove any residual information accessible to Eve.
Finally, concrete proposals for practical implementation remain an important aspect for future work.

%---------------------------------------------------------------------
\section{Conclusions}\label{sec:conclusions}
%---------------------------------------------------------------------

\noindent
In this paper, we propose a novel one-way \ac{qkd} protocol based on quantum walks over a hypercube topology.
We also provide the first Qiskit implementation and simulation of one-way \ac{qkd} using quantum walks on both circle and hypercube graphs, designing two versions: one exploiting the circle's symmetry and the other leveraging the hypercube's high connectivity.
Our prepare-and-measure framework, built entirely in Qiskit, supports arbitrary topologies and built-in noise models (e.g., depolarizing, amplitude-phase damping), enabling a detailed performance comparison under realistic conditions.
To ensure flexibility and reproducibility, all code and results are publicly available on GitHub~\cite{github_repo}.
Our findings confirm recent results by Yu et al.~\cite{qkd_qudits} that high-dimensional systems are more noise-resilient than qubit-based schemes.
In particular, quantum-walk states in a hypercube show superior noise robustness and security compared to the state-of-the-art, as detailed in Sections~\ref{sec:security_evaluation} and~\ref{sec:noise_resistance_analysis}, despite the limitations discussed in Section~\ref{sec:discussion}.
While practical implementation poses challenges (especially in realizing high-dimensional walks) recent advances in quantum walks and high-dimensional state engineering suggest feasibility in current photonic and simulator platforms.
Crucially, robustness does not rely solely on large dimensions, indicating that experimental realization may already be feasible to implement.
Overall, our results highlight quantum walks, especially on hypercubes, as a powerful and flexible approach to secure quantum communication.

%---------------------------------------------------------------------
% Appendix
%---------------------------------------------------------------------
\appendices

\section{One-way QKD and entanglement-based schemes}\label{apn:qkd_protocol_schemes}

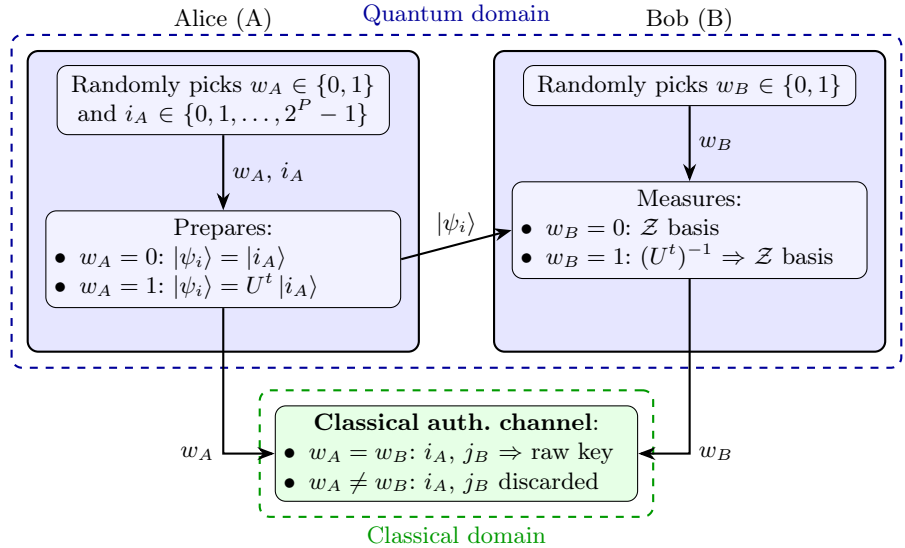
\begin{figure}[!ht]
    \centering
    \scalebox{0.72}{
    \begin{tikzpicture}[transform shape, node distance=1cm,
    every node/.style={align=center}, 
    bigbox/.style={draw, rectangle, thick, minimum width=5.2cm, minimum height=4cm, rounded corners, fill=blue!10},
    box/.style={draw, rectangle, rounded corners, text width=4.5cm, minimum height=0.3cm, text centered, fill=blue!5}, 
    smallbox/.style={draw, rectangle, rounded corners, text width=4.2cm, minimum height=0.4cm, text centered, fill=blue!5}, 
    sharedbox/.style={draw, rectangle, rounded corners, text width=4.6cm, minimum height=1cm, text centered, fill=green!10}, 
    arrow/.style={-{Stealth}, thick}]
    % Large boxes for Alice and Bob
    \node[bigbox, label={[yshift=0.12cm]Alice (A)}] (alice_big) at (0, 0) {};        
    \node[bigbox, label={[yshift=0.12cm]Bob (B)}] (bob_big) at (6.2cm, 0) {};
    % Alice's internal nodes
    \node[smallbox, anchor=north] (alice_pick) at ([yshift=-0.2cm]alice_big.north) {Randomly $w_A \in \{0, 1\}$, $i_A \in \{0, 1, \dots, 2^{P}-1\}$};
    \node[box, below=of alice_pick] (alice_prepare) {
    Prepares $\ket{\psi_i}$:
    \begin{itemize}[leftmargin=*, label=\textbullet, noitemsep, topsep=0pt, partopsep=0pt]
        \item $w_A = 0$: $\ket{i_A}$
        \item $w_A = 1$: $U^t \cdot (I \otimes F) \ket{i_A}$
    \end{itemize}};
    % Bob's internal nodes
    \node[smallbox, anchor=north] (bob_pick) at ([yshift=-0.2cm]bob_big.north) {Randomly $w_B \in \{0, 1\}$};
    \node[box, below=of bob_pick] (bob_measure) {Measures $\ket{\psi_i}$:
    \begin{itemize}[leftmargin=*, label=\textbullet, noitemsep, topsep=0pt, partopsep=0pt]
        \item $w_B = 0$: $\mathcal{Z}$ basis
        \item $w_B = 1$: $(U^t)^{-1}$ $\Rightarrow$ $\mathcal{Z}$ basis
    \end{itemize}};
    % Quantum operations box
    \node[draw=blue!60!black, thick, dashed, rounded corners, fit=(alice_big)(bob_big), inner sep=0.2cm,
    label={[blue!60!black]above:Quantum domain}
    ] {};;
    % Shared decision box
    \path (alice_big.south) -- (bob_big.south) coordinate[midway] (midpoint);
    \node[sharedbox, below=0.7cm of midpoint] (decision)
    {\textbf{Classical auth. channel}:
    \begin{itemize}[leftmargin=*, label=\textbullet, noitemsep, topsep=0pt, partopsep=0pt]
        \item $w_A = w_B$: $i_A$, $j_B$ raw key
        \item $w_A \neq w_B$: $i_A$, $j_B$ discarded
    \end{itemize}};
    % Classical domain box
    \node[draw=green!60!black, thick, dashed, rounded corners, fit=(decision), inner sep=0.2cm, label={[green!60!black]below:Classical domain}] {};
    % Arrows between nodes
    \draw[arrow] (alice_pick.south) -- (alice_prepare.north) node[midway, right] {$w_A$, $i_A$};
    \draw[arrow] (bob_pick.south) -- (bob_measure.north) node[midway, right] {$w_B$};
    \draw[arrow] (alice_prepare.east) -- (bob_measure.west) node[midway, above] {$\ket{\psi_i}$};
    \draw[arrow] (alice_prepare.south) -- ++(0, -1) |-(decision.west) node[midway, left] {$w_A$};
    \draw[arrow] (bob_measure.south) -- ++(0, -1) |-(decision.east) node[midway, right] {$w_B$};
    \end{tikzpicture}
    }
    \caption{Schematic representation of the proposed one-way hypercube-based \ac{qkd} protocol. The arrow from Alice to Bob represents the transmission over the public quantum channel of the state $\ket{\psi_i}$ prepared by Alice, based on $w_A$ and $i_A$. After the quantum communication, a classical authenticated channel is used to compare the basis choices (i.e., $w_A$, $w_B$) and extract the shared key.}
    \label{fig:one_way_protocol_h}
\end{figure}

\noindent
In this section, we present the proposed one-way \ac{qkd} protocol based on quantum walks over a hypercube topology and its entanglement-based variant to illustrate its security.
Figure~\ref{fig:one_way_protocol_h} shows the main steps: Alice randomly selects a basis $w_A$ and state index $i_A$, prepares the corresponding quantum state $\ket{\psi_i}$, and sends it to Bob. Bob randomly chooses a measurement basis $w_B$ and measures the received state.
Using a classical authenticated channel, they compare basis choices, keeping outcomes only when $w_A = w_B$ to form the raw key.

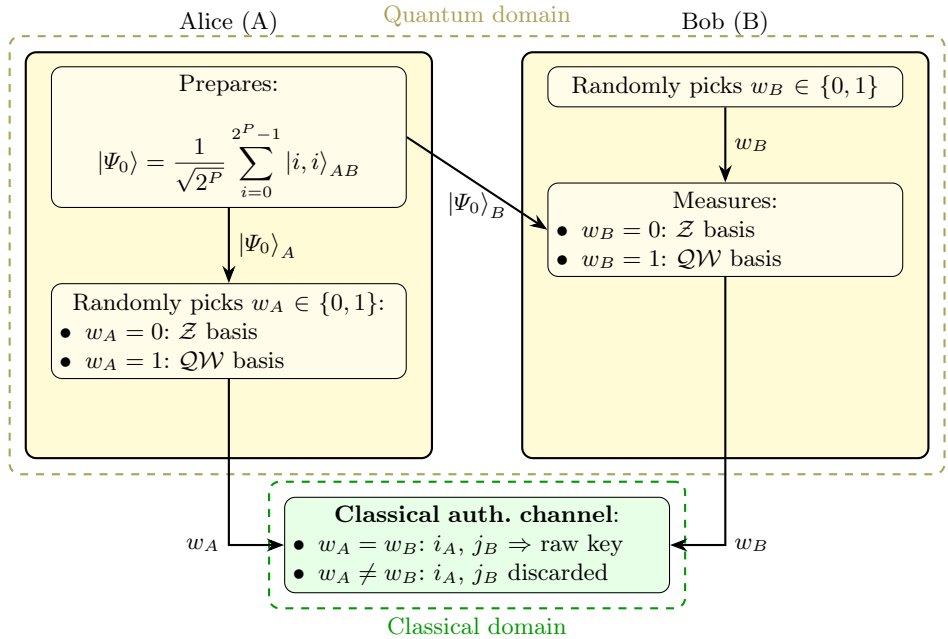
\begin{figure}[!ht]
    \centering
    \scalebox{0.68}{
    \begin{tikzpicture}[transform shape, node distance=1cm,
        every node/.style={align=center}, bigbox/.style={draw, rectangle, thick, minimum width=5.4cm, minimum height=5.4cm, fill=yellow!20, rounded corners}, box/.style={draw, rectangle, rounded corners, text width=4.5cm, minimum height=0.3cm, text centered, fill=yellow!10}, smallbox/.style={draw, rectangle, rounded corners, text width=4.5cm, minimum height=0.4cm, text centered, fill=yellow!10}, sharedbox/.style={draw, rectangle, rounded corners, text width=4.9cm, minimum height=1cm, fill=green!10, text centered}, arrow/.style={-{Stealth}, thick}]
        % Large boxes for Alice and Bob
        \node[bigbox, label={[yshift=0.1cm]Alice (A)}] (alice_big) at (0, 0) {};
        \node[bigbox, label={[yshift=0.1cm]Bob (B)}] (bob_big) at (6.6cm, 0) {};
        % Alice's internal nodes
        \node[box, anchor=north] (alice_prepare) at ([yshift=-0.2cm]alice_big.north) {
        Prepares: \[\ket{\Psi_0} = \frac{1}{\sqrt{2^P}} \sum_{i=0}^{2^P-1} \ket{i, i}_{AB}\]
        };
        \node[box, below=of alice_prepare] (alice_measure) {
        Randomly picks $w_A \in \{0, 1\}$:
        \begin{itemize}[leftmargin=*, label=\textbullet, noitemsep, topsep=0pt, partopsep=0pt]
            \item $w_A = 0$: $\mathcal{Z}$ basis
            \item $w_A = 1$: $\mathcal{QW}$ basis
        \end{itemize}
        };
        % Bob's internal nodes
        \node[smallbox, anchor=north] (bob_pick) at ([yshift=-0.2cm]bob_big.north) {Randomly picks $w_B \in \{0, 1\}$};
        \node[box, below=of bob_pick] (bob_measure) {
        Measures:
        \begin{itemize}[leftmargin=*, label=\textbullet, noitemsep, topsep=0pt, partopsep=0pt]
            \item $w_B = 0$: $\mathcal{Z}$ basis
            \item $w_B = 1$: $\mathcal{QW}$ basis
        \end{itemize}
        };
        % Quantum operations box
        \node[draw=yellow!60!black, thick, dashed, rounded corners, fit=(alice_big)(bob_big), inner sep=0.2cm, label={[yellow!60!black]above:Quantum domain}] {};;
        % Shared decision box centered
        \path (alice_big.south) -- (bob_big.south) coordinate[midway] (midpoint);
        \node[sharedbox, below=0.5cm of midpoint] (decision) {
        \textbf{Classical auth. channel}:
        \begin{itemize}[leftmargin=*, label=\textbullet, noitemsep, topsep=0pt, partopsep=0pt]
            \item $w_A = w_B$: $i_A$, $j_B$ $\Rightarrow$ raw key
            \item $w_A \neq w_B$: $i_A$, $j_B$ discarded
        \end{itemize}
        };
        % Classical operations box
        \node[draw=green!60!black, thick, dashed, rounded corners, fit=(decision), inner sep=0.2cm, label={[green!60!black]below:Classical domain}] {};;
        % Arrows between nodes
        \draw[arrow] (alice_prepare.south) -- (alice_measure.north) node[midway, right] {$\ket{\Psi_0}_{A}$};
        \draw[arrow] (bob_pick.south) -- (bob_measure.north) node[midway, right] {$w_B$};
        \draw[arrow] (alice_prepare.east) --(bob_measure.west) node[midway, below] {$\ket{\Psi_0}_{B}$};
        \draw[arrow] (alice_measure.south) -- ++(0, -1) |-(decision.west) node[midway, left] {$w_A$};
        \draw[arrow] (bob_measure.south) -- ++(0, -1) |-(decision.east) node[midway, right] {$w_B$};
    \end{tikzpicture}
    }
    \caption{Schematic representation of the entanglement-based one-way \ac{qkd} protocol on a hypercube topology (see Figure~\ref{fig:one_way_protocol_h}). Alice prepares the entangled state $\ket{\Psi_0}$, retains one half, and sends the second half to Bob. Both parties perform measurements in a randomly selected basis ($\mathcal{Z}$ or $\mathcal{QW}$), discarding outcomes when their basis choices differ.}
    \label{fig:ent_based_one_way_protocol_h}
\end{figure}

\noindent
To demonstrate the protocol's security, Figure~\ref{fig:ent_based_one_way_protocol_h} depicts the entanglement-based version.
Alice prepares a maximally entangled state $\ket{\Psi_0}$, keeps one half, and sends the other to Bob.
Both parties randomly measure in the $\mathcal{Z}$ or $\mathcal{QW}$ basis, discarding outcomes when bases differ.

\noindent
Together, these representations provide a complete view of the hypercube-based \ac{qkd} protocol, showing both practical prepare-and-measure operations and the entanglement-based scheme that underpins its security.

\section{Qiskit implementation of quantum walks}\label{apn:qiskit_implementation}

\begin{figure}[!ht]
    \centering
    \includegraphics[width=\columnwidth]{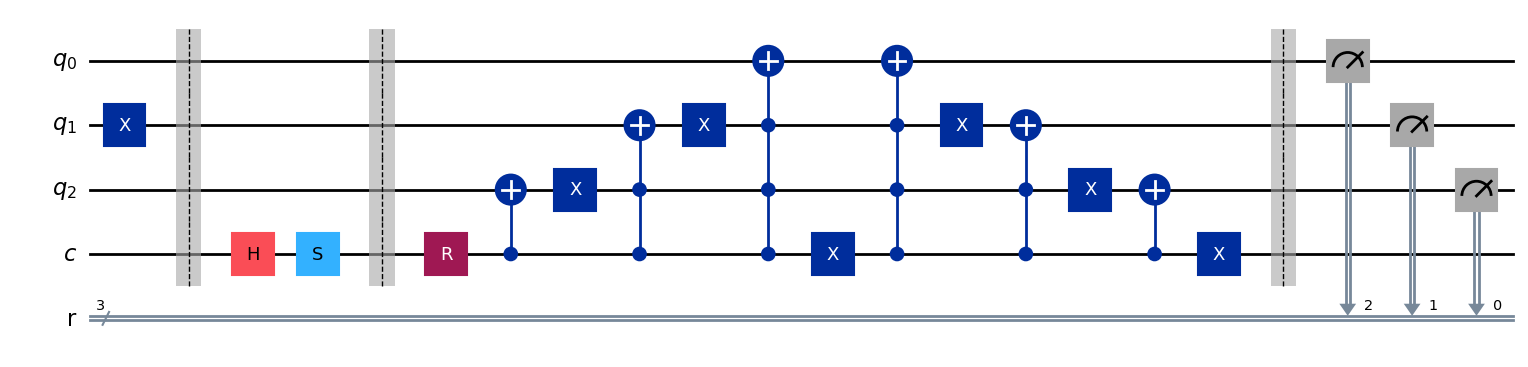}
    \caption{Qiskit complete implementation of a circle-based quantum walk: $P = 3$, $t = 1$, $\ket{\psi_0} = \ket{2} = \ket{(10)_2}$, $F = Y$, $\phi = 0$, $\theta = \pi/4$.}
    \label{fig:qiskit_qw_circle_sample}
\end{figure}

\begin{figure}[!ht]
    \centering
    \includegraphics[width=\columnwidth]{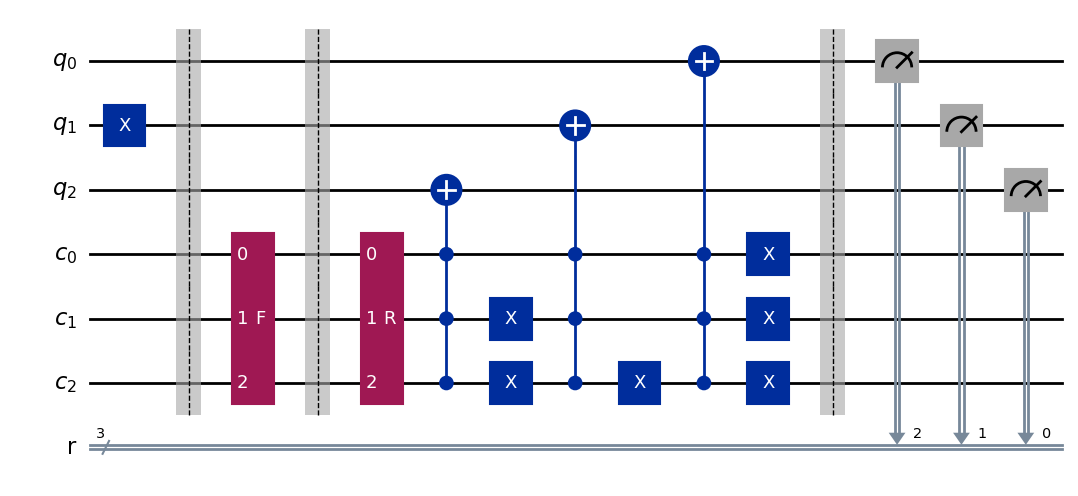}
    \caption{Qiskit complete implementation of a hypercube-based quantum walk (generic coin): $P = 3$, $t = 1$, $\ket{\psi_0} = \ket{2} = \ket{(10)_2}$, $F = Y$, $\phi = 0$, $\theta = \pi/4$.}
    \label{fig:qiskit_qw_hypercube_sample_c}
\end{figure}

\noindent
In this section, we present the Qiskit implementation of our quantum walk schemes, which serve as realistic circuit-level representations of the models described in the main work. 
Although achieving a nearly uniform distribution of states requires a large number of steps (i.e., $t \gg 1$), we restrict our discussion to the single-step case ($t = 1$) for clarity, since visualizing longer evolutions quickly becomes impractical.
The presented circuits highlight how the position and coin spaces are encoded in qubit registers, and how the walk operators is built from elementary quantum gates. 
Specifically, Figures~\ref{fig:qiskit_qw_circle_sample} and~\ref{fig:qiskit_qw_hypercube_sample_c} show the structures corresponding to the circle-based walk, and the hypercube walk with a generic rotation coin, respectively. 

\noindent
These implementations not only provide an illustrative reference for the theoretical framework but also demonstrate their feasibility on near-term quantum devices.

%---------------------------------------------------------------------
% Acknowledgment
%---------------------------------------------------------------------
\section*{Acknowledgment}

\noindent
The authors would like to thank the University of Padova and the Institute of Physics of the Slovak Academy of Sciences for supporting this research.
We also acknowledge the valuable contributions and discussions provided by colleagues from the University of Lisbon and the University of Connecticut during the early stages of this work.
This research was supported in part by the European Union through the Connecting Europe Facility (CEF Digital) under Grant Agreement No. 101249538 (``Implement QCI Network in Central Europe - CEQCI'').

%---------------------------------------------------------------------
% Bibliography
%---------------------------------------------------------------------
\bibliographystyle{IEEEtran}
\bibliography{references}

\end{document}